\documentclass[aps, twocolumn,noshowpacs,preprintnumbers,amsmath,amssymb]{revtex4}

\usepackage{verbatim}

\usepackage{chemstyle} 
\usepackage{graphicx}
\usepackage{dcolumn}
\usepackage{bm}
\usepackage{wrapfig}

\begin{document}

\title{CdS Nanoparticles Capped with 1-Substituted 5-Thiotetrazoles: Synthesis, Characterization, and Thermolysis of the Surfactant}
\author{Sergei V. Voitekhovich}
\affiliation{Institute of Physical Chemistry, University of Hamburg, D - 20146 Hamburg, Germany}
\affiliation{Research Institute for Physical Chemical Problems, Belarussian State UniVersity, 220050 Minsk, Belarus}
\author{Dmitri V. Talapin}
\affiliation{Department of Chemistry, The University of Chicago, Chicago, Illinois 60637, USA}
\author{Christian Klinke}
\author{Andreas Kornowski}
\author{Horst Weller}
\affiliation{Institute of Physical Chemistry, University of Hamburg, D - 20146 Hamburg, Germany }

\maketitle

During the past decade semiconductors, metals, and magnetic materials, synthesized in form of nanoparticles (NPs), have attracted considerable attention due to their novel physical and chemical properties and their large potential for various industrial applications. Semiconductor nanomaterials are promising candidates as building blocks of future electronic, optoelectronic, and photonic devices \cite{1}. Moreover, they can be widely employed in biological and medical applications \cite{2}. NPs with tunable sizes and different shapes have been synthesized by the methods of colloidal chemistry \cite{1,3}. Generally, colloidally synthesized NPs comprise a crystalline core surrounded with a layer of surfactant molecules. The surfactant molecules (also referred to as "stabilizing agents" and "ligands") bind to the NP surface immediately after its nucleation and prevent them from fast growth and coagulation. The shell of stabilizing agents attached to the NP surface
also provides solubility of NPs in a desired solvent which is very important for processing NPs \cite{4}. On the other hand, the surfactant molecules can significantly change the NPs' properties. For example, they can block catalytic surface sites, leading to a dramatic decrease of the catalytic activity of transition metal NPs \cite{5}. Insulating shells around semiconducting NPs lead to very low conductivities of NP films limiting the use of NPs in electronic devices and solar cells \cite{6}. Typically, alkyl phosphines and phosphine oxides, long chain alkylamines, and carboxylic and phosphonic acids are used as stabilizing agents in NP synthesis. The attempts of complete removal of these molecules by heating the NP layers in vacuum were unsuccessful because of partial destruction and carbonization of surfactant molecules at high temperatures \cite{7}. There are several reports on chemical removal of stabilizing agents, for example, by treating semiconducting
NPs with sodium hydroxide8 or hydrazine \cite{9}. These techniques, although improving the electronic properties of semiconducting NPs, do not allow complete removal of stabilizing agents and can only be applied to a limited range of materials and surfactants.

\begin{scheme}[ht]
\includegraphics[width=0.9\textwidth]{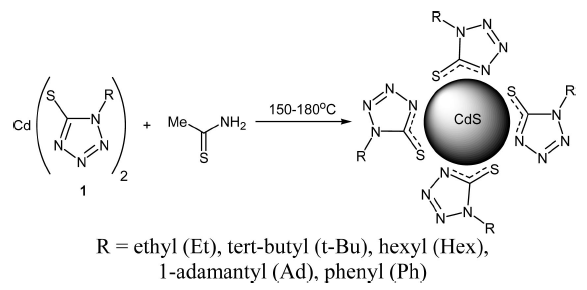}
\caption{Solution-Phase Synthesis of CdS NPs Capped with 1-R-5-Thiotetrazoles}
\end{scheme}

We propose a new type of surfactants, namely, tetrazole derivatives which can be controllably removed from the NP surface. Tetrazoles are a peculiar class of heterocyclic compounds. The presence of four nitrogen atoms in the tetrazole ring determines their interesting physical and chemical properties. Tetrazoles show high thermal stability below 200$^{\circ}$C while decomposing at higher temperature with formation of gaseous products and no or very little solid residue \cite{10}. Moreover, the tetrazole group is known as an important ligand in coordination chemistry. Its donor nitrogen atoms can bind to various metal ions leading to stable
complexes with diverse coordination modes of the heteroring \cite{11}.

\begin{figure*}[ht!]
\begin{center}
\includegraphics[width=0.9\textwidth]{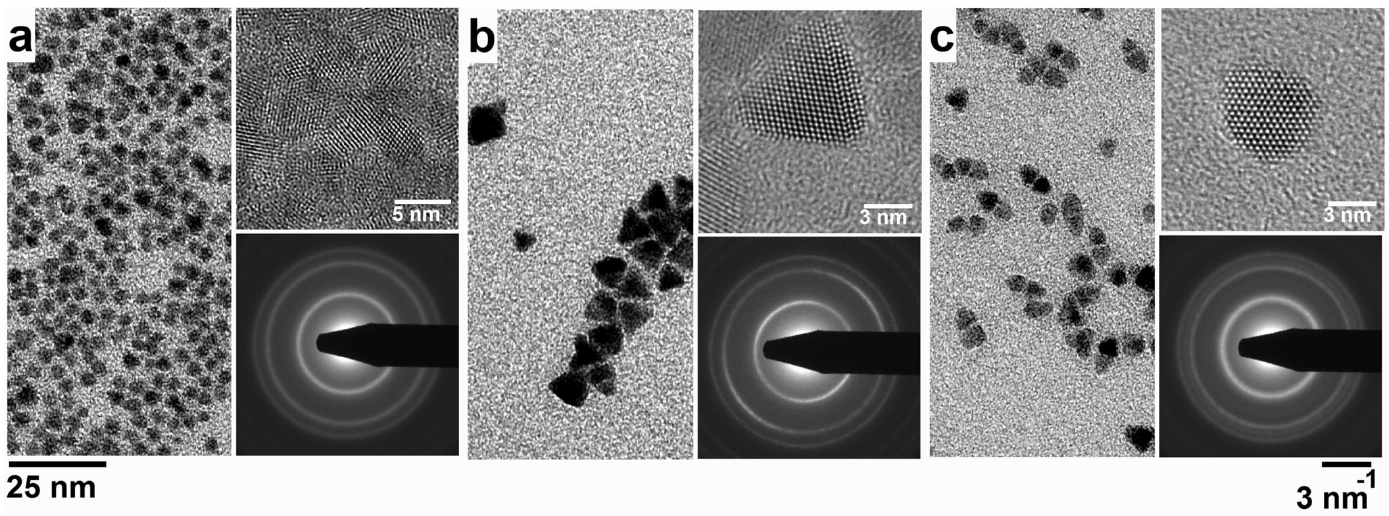}
\caption{\it TEM images of CdS NPs as-synthesized by solution-phase approach using 1-R-5-thiotetrazolates \textbf{1} (a) R = Et, (b) R = Hex, and (c) R = \textit{t}-Bu.}
\end{center}
\end{figure*}

Here, we report the synthesis of CdS NPs capped with 1-R-5-thiotetrazoles using two different synthetic schemes: solution-phase (Scheme 1) and solventless single precursor approaches. In both synthetic procedures cadmium thiotetrazolates \textbf{1} were used as cadmium precursors and sources of surfactant.

\begin{figure}[!h]
\begin{center}
\includegraphics[width=0.9\textwidth]{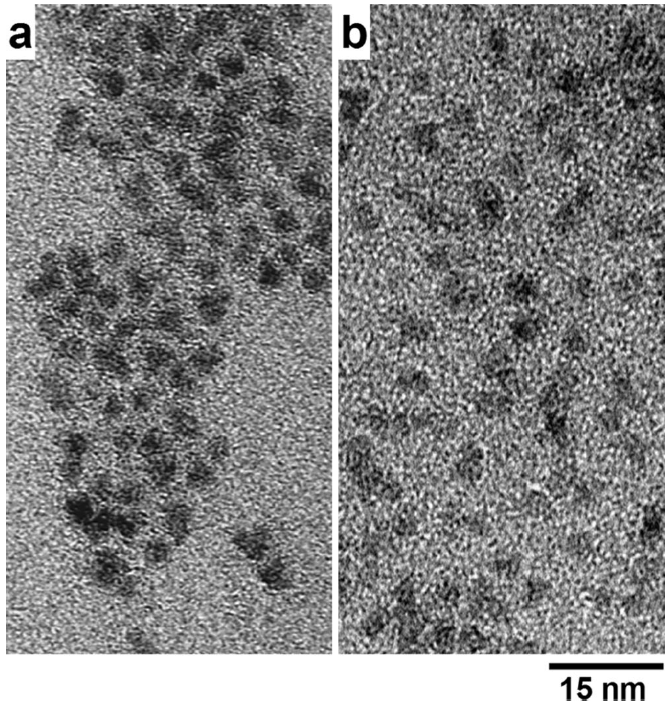}
\caption{\it TEM images of CdS NPs as-synthesized by solventless thermolysis of cadmium 1-R-5-thiotetrazolates \textbf{1} with (a) R = Ad, (b) R = Et.}
\end{center}
\end{figure}

A typical solution-phase synthesis yielding CdS NPs capped with thiotetrazoles was carried out under nitrogen using standard Schlenk line technique. The mixture of salt \textbf{1} (0.25 mmol) and 1,2-dichlorobenzene (25 mL) was purged with nitrogen for 15 min and heated to 150 or 180$^{\circ}$C. A solution of thioacetamide (23 mg, 0.3 mmol) in DMF (0.2 mL) was injected into the dichlorobenzene solution, and the reaction mixture was kept at constant temperature for 1.5 h. Finally, the solvent was evaporated under reduced pressure, and the residue was washed with hexane or acetonitrile, dissolved in chloroform, and filtered through a PTFE 0.2 $\mu$m filter. Figure 1a shows transmission electron microscopy (TEM) images of monodisperse 3.7 $\pm$ 0.3 nm diameter CdS NPs synthesized from cadmium 1-ethyl-5-thiotetrazolate at 150$^{\circ}$C. Under heating at 180$^{\circ}$C CdS NPs with different shapes and sizes are formed depending on the nature of
substituent R (Table 1). Whereas with R = Et or Ad spherical NPs were obtained, R = Hex yields tetrahedral NPs with $\sim$8 nm edge (Figure 1b). In the case of R = \textit{t}-Bu both spherical and tetrahedral particles were obtained (Figure 1c). Salt \textbf{2} (R = Ph), finally, gives CdS NPs, which coalesce
into polycrystalline aggregates.

\begin{table*}
\caption{Summary of the CdS NPs obtained using various cadmium precursors 1}
    \begin{tabular}{cccc} 
				\hline
        R    & Size [nm] & Morphology & NP structure \\ \hline
				Et &  4.7 $\pm$ 0.6 & spherical & zinc blende \\ 
				Hex & 8.5 $\pm$ 0.6 & tetrahedral & zinc blende \\
				t-Bu & 4.8 $\pm$ 0.5 & tetrahedral and spherical & wurtzite \\ 
				Ad & 2.9 $\pm$ 0.3 & spherical & zinc blende \\ 
				Ph &   & polycrystalline aggregates & zinc blende \\ 
    \hline
    \end{tabular}
\end{table*}

Solventless synthesis of CdS NPs is based on thermal decomposition of single precursors \textbf{1} (R = Et, Ad). Synthesis yielding spherical 2.5 -- 3 nm CdS NPs (Figure 2) is described as follows. The salt \textbf{1} ($\sim$10 mg for R = Et or $\sim$20 mg for R = Ad) was heated under a flow of nitrogen to 250$^{\circ}$C with a rate of 5$^{\circ}$C/min and kept for 1 h (R = Et) or 2 h (R = Ad) at 250$^{\circ}$C. The obtained yellow solid was dissolved in chloroform and filtered. For purification the NPs were precipitated with methanol and redispersed in chloroform. To our knowledge this is the first example of a solventless single precursor synthesis of CdS NPs. To date the thermolytic method has proven to be successful in producing other nanomaterials, such as Cu$_{2}$S \cite{12}, NiS \cite{13}, PbS \cite{14}, Bi$_{2}$S$_{3}$ \cite{15}, Fe$_{3}$O$_{4}$ \cite{16,17}, Ag \cite{17} and Bi \cite{18} using metal alkylthiolates and oleates as precursors.

All above-mentioned NPs belong to cubic phase CdS (zinc blende structure) except the NPs synthesized from 1-tertbutyl-5-thiotetrazole whose X-ray diffraction (XRD) pattern can be consistently indexed as hexagonal wurtzite type CdS phase. The obtained NPs are stable in solution under ambient conditions for a few months. Their solubility strongly depends on the nature of substitutent R. NPs with R = Et are very soluble in acetonitrile, slightly in chloroform, and not soluble in hexane. In contrast when R = Ad, the NPs are soluble in hexane but not soluble in acetonitrile. The absorption spectra of the obtained NPs show excitonic features characteristic to CdS NPs. Moreover, CdS NPs synthesized by the solution-phase approach exhibited weak photoluminescence. Their emission spectra ($\lambda _{ex}$ = 450 nm) showed a broad photoluminescence band with maxima located at about 615 -- 705 nm depending on substitutent R of thiotetrazolate \textbf{1}.

\begin{figure}[!h]
\begin{center}
\includegraphics[width=0.9\textwidth]{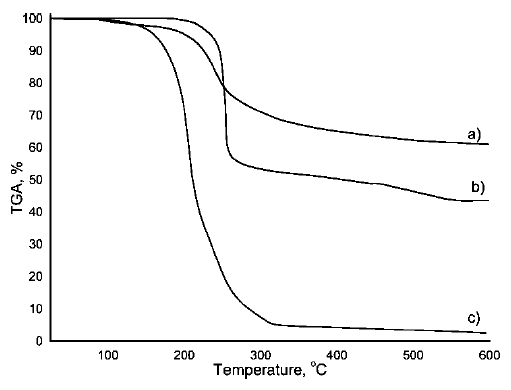}
\caption{\it TG curves of (a) CdS NPs stabilized with 1-ethyl-5-thiotetrazole, (b) cadmium 1-ethyl-5-thiotetrazolate, and (c) 1-ethyl-5-thiotetrazole.}
\end{center}
\end{figure}

The presence of thiotetrazolate anions at the surface of NPs is confirmed by IR spectroscopy. All samples show the characteristic tetrazole bands. In particular, the tetrazole ring stretching vibrations are identified in regions 1020 -- 1160 cm$^{-1}$ \cite{19}. A more careful inspection of the spectra shows, however, that relative intensities and exact peak positions of the tetrazole bands vary for the different samples. This finding is understood in terms of various coordination modes
of thiotetrazolate moiety at the surface of NPs. Binding can occur via sulfur or nitrogen or by both atoms simultaneously \cite{11}.

\begin{scheme}[ht]
\includegraphics[width=0.9\textwidth]{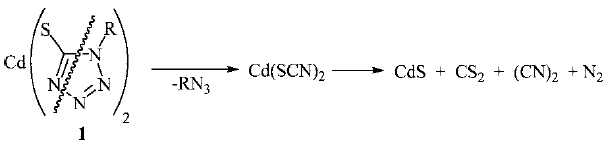}
\caption{Thermolysis of Cadmium 1-R-5-Thiotetrazolates}
\end{scheme}

Figure 3 shows the thermogravimetric (TG) analysis curves of 1-ethyl-5-thiotetrazole, the corresponding cadmium salt \textbf{1} (R = Et), and the respective NPs obtained by solution phase synthesis. Whereas the free ligand decomposes at $\sim$150$^{\circ}$C, both the cadmium salt and the NPs are more stable and decompose at $\sim$240$^{\circ}$C. This behavior is usual for anionic derivatives of tetrazoles and due to higher aromacity of the tetrazolate ring in comparison with neutral tetrazole ring \cite{11}. After heating to 600$^{\circ}$C under nitrogen atmosphere, salt \textbf{1} (R = Et) yields an orange solid residue. According to XRD this residue consists of wurtzite CdS. Moreover, the residual mass after the decomposition correlates with the calculated CdS content in salts. A possible pathway of
the thermal decomposition of cadmium thiotetrazolates is depicted in Scheme 2.

The salt \textbf{1} undergoes ring fragmentation under formation of gaseous organic azide and solid cadmium thiocyanate. The intermediate formation of thiocyanate is confirmed by presence of $\nu$(CN) bands of this moiety in the region 2070 -- 2170 cm$^{-1}$ of IR spectra of CdS NPs obtained by thermal decomposition of cadmium 1-ethyl-5-thiotetrazolate \cite{20}. In a further step cadmium thiocyanate decomposes leaving only the appropriate sulfides as a solid phase \cite{21}. We can assume the same method of thermal decomposition of thiotetazolates in the case of NPs and salts \textbf{1}. It means that theoretically only sulfur atoms remain on the surface of NPs after decomposition of thiotetrazolate. Actually, heating of powder CdS NPs capped with 1-ethyl-5-thiotetrazolate at 250-270$^{\circ}$C for 1.5 -- 2 h under nitrogen leads to a solid product which is insoluble in organic solvents. Moreover, the IR spectrum of that solid does not show any absorption in region 600 -- 4000 cm$^{-1}$ which confirms the complete decomposition of the tetrazoles.

In summary, a facile and simple solution-phase and solventless single precursor synthesis of thiotetrazole capped CdS NPs was described for the first time. The shape of the obtained NPs depends on the nature of the substitutent in the surfactant. Conditions for thermal decomposition of the surfactant, namely, 1-substituted 5-thiotetrazolate anions, on the surface of NPs were found. Finally, thiotetrazoles are proposed as a new type of surfactants for colloidal synthesis of high quality NPs. These surfactants are characterized by low carbon content and thermally induced decomposition under formation of gaseous products. Further design of tetrazole capped NPs and study of their properties can provide a solution to the long-standing problem of making efficient catalysts and electronic materials using colloidal
chemistry techniques.

\section*{Acknowledgment}

We gratefully acknowledge the support of the Alexander von Humboldt Foundation (Research Fellowship of Dr. S. Voitekhovich).

\clearpage

\end{document}